\documentclass[useAMS,usenatbib]{mn2e}
\usepackage[dvips]{graphicx}
\usepackage{times}
\usepackage{fixltx2e}
\newcommand{\aap}{A\&A}

\newcommand{\mnras}{MNRAS}

\newcommand{\apj}{ApJ}
\newcommand{\apjl}{ApJ}
\newcommand{\apjs}{ApJS}
\newcommand{\pasj}{PASJ}

\newcommand{\aj}{AJ}

\newcommand{\msun}{M$_{\odot}$}

\title[Multiple Populations in Magellanic Cloud Clusters]
      {The dynamical origin of multiple populations in intermediate-age clusters in the Magellanic Clouds}
\author[J. Hong, et al.]
{Jongsuk Hong$^1$\thanks{E-mail: jongsuk.hong@pku.edu.cn (JH);}, Richard de Grijs$^{1,2,3}$, 
Abbas Askar$^4$, Peter Berczik$^{5,6}$, Chengyuan Li$^7$\\
\newauthor Long Wang$^8$, Licai Deng$^9$, M.B.N. Kouwenhoven$^{10}$, Mirek Giersz$^4$ and Rainer Spurzem$^{6,1,11}$ \\
$^1$Kavli Institute for Astronomy and Astrophysics, Peking University, Yi He Yuan Lu 5, HaiDian District, Beijing 100871, China\\
$^2$Department of Astronomy, School of Physics, Peking University, Yi He Yuan Lu 5, HaiDian District, Beijing 100871, China\\
$^3$International Space Science Institute -- Beijing, 1 Nanertiao, Zhongguancun, Hai Dian District, Beijing, 100190, China\\
$^4$Nicolaus Copernicus Astronomical Centre, Polish Academy of Sciences, ul. Bartycka 18, 00-716 Warsaw, Poland\\
$^5$Main Astronomical Observatory, National Academy of Sciences of Ukraine, 27 Akademika Zabolotnoho St., 03143 Kyiv, Ukraine\\
$^6$Key Laboratory for Computational Astrophysics, National Astronomical Observatories, Chinese Academy of Sciences, 20A Datun Road,\\ Chaoyang District, Beijing 100012, China\\
$^7$Department of Physics and Astronomy, Macquarie University, Sydney, NSW 2109, Australia\\
$^8$RIKEN Advanced Institute for Computational Science, 7-1-26 Minatojima-minami-machi, Chuo-ku, Kobe, Hyogo, Japan\\
$^9$Key Laboratory for Optical Astronomy, National Astronomical Observatories, Chinese Academy of Sciences, 20A Datun Road,\\ Chaoyang District, Beijing 100012, China\\
$^{10}$Department of Mathematical Sciences, Xi’an Jiaotong-Liverpool University, 111 Ren’ai Rd., \\ Suzhou Dushu Lake Science
and Education Innovation District, Suzhou Industrial Park, Suzhou 215123, P.R. China\\
$^{11}$Astronomisches  Rechen-Institut,  Zentrum  f\"ur  Astronomie,  University  of  Heidelberg, M\"onchhofstrasse 12-14, 69120 Heidelberg, Germany
}
\begin{document}

\date{Accepted 2017 July 26. Received 2017 July 26; in original form 2017 May 29}
\pagerange{\pageref{firstpage}--\pageref{lastpage}} \pubyear{2017}
\maketitle
\label{firstpage}

\begin{abstract}
Numerical simulations were carried out
to study the origin of multiple stellar populations in the intermediate-age clusters NGC 411 and NGC 1806 in the Magellanic Clouds.
We performed {\sc nbody6++} simulations based on two different formation scenarios, an ad hoc formation model where second-generation (SG) stars are formed inside a cluster of first-generation (FG) stars using the gas accumulated from the external intergalactic medium
and a minor merger model of unequal mass ($M_{\rm SG}/M_{\rm FG} \sim 5$--10\%) clusters with an age difference of a few hundred million years.
We compared our results such as the radial profile of the SG-to-FG number ratio with observations on the assumption that the 
SG stars in the observations are composed of cluster members, and confirmed that
both the ad hoc formation and merger scenarios reproduce the observed radial trend of the SG-to-FG number ratio which shows less centrally concentrated SG than FG stars.
It is difficult to constrain the formation scenario for the multiple populations by only using the spatial distribution of the SG stars.
SG stars originating from the merger scenario show a significant velocity anisotropy and rotational features compared to those from the ad hoc formation scenario.
Thus, observations aimed at kinematic properties like velocity anisotropy or rotational velocities for SG stars should be obtained 
to better understand the formation of the multiple populations in these clusters. 
This is, however, beyond current instrumentation capabilities.
\end{abstract}

\section{INTRODUCTION}
Star clusters are stellar systems composed of large numbers of stars bounded by their self-gravity which are among the common building blocks of galaxies. 
Star clusters are considered ideal laboratories to study stellar evolution, since stars in clusters 
are usually assumed to be coeval and share the same chemical properties. However, recent observations 
have revealed that the star-formation history inside clusters is not-so-simple.
Observational evidence that globular clusters host multiple stellar populations has come from both spectroscopic studies, which revealed that cluster stars are often characterized by
different chemical abundances (the so-called Na-O anti-correlation; see e.g. Gratton et al.
2012, and references therein) and photometric studies which present distinguishable sequences of stars in colour--magnitude diagrams (CMDs) at different stages of 
stellar evolution (Lee et al. 1999; Bedin et al. 2004; Siegel et al. 2007; Piotto et al. 2007, 2015; 
Milone et al. 2008, 2010, 2012; Bellini et al. 2013; Li et al. 2014). 

Different sources of gas for the second-generation 
(SG) star formation have been suggested (see e.g. Ventura et al. 2001; Decressin et al. 2007; Bastian et al. 2013; 
note that the term `generation' does not apply to the model proposed by Bastian et al. 2013 since in that model there are no separate star formation episodes)
and a number of studies have addressed some of the issues concerning the origin of the observed abundance 
patterns and the formation and dynamical evolution of multiple-population clusters (see e.g. D'Ercole et al. 2008). 
Although no consensus has yet been reached on this fundamental issue, all models proposed so far agree that 
SG stars should form in the central regions of a more diffuse first-generation (FG) system 
(e.g. D'Ercole et al. 2008); a number of observational studies have confirmed this prediction and found 
several clusters in which SG stars are more spatially concentrated than FG stars (e.g., Sollima et al. 2007; 
Bellini et al. 2009; Lardo et al. 2011; Milone et al. 2012; Beccari et al. 2013; Cordero et al. 2014; Kucinskas et al. 2014; Li et al. 2014; 
Simioni et al. 2016). Clusters not showing any difference in radial distribution of the two populations could 
also have had a radial discrepancy in the past, since some theoretical studies focusing on the long-term 
evolution of multiple-population clusters have predicted that spatial mixing of different-generation stars will 
be achieved when the cluster loses a significant fraction of its mass (Vesperini et al. 2013; 
Miholics et al. 2015). Multiple stellar populations also show kinematic discrepancies; 
recent {\sl Hubble Space Telescope} proper motion observations have revealed that younger-generation 
stars tend to show more radial anisotropy than older-generation stars (Richer et al. 2013; Bellini et al. 2015).
Furthermore, Cordero et al. (2017) have found differential rotation among multiple populations in  
M13 using radial velocity measurements. 

The Large and Small Magellanic Clouds (LMC/SMC) are gas-rich galaxies interacting with the Milky Way and showing 
a vigorous star-formation history until recently (Rubele et al. 2012). The LMC/SMC host numerous young ($<$100 Myr old) and 
intermediate-age star clusters, where there are no Milky Way counterparts, allowing to 
examine the presence of multiple populations in different environments. Based on accurate photometric observations over the last decade, 
there have been several studies that found multiple stellar populations in some young/intermediate-age massive clusters in the Magellanic Clouds 
based on their extended main-sequence turn-offs (eMSTOs) (e.g., Mackey et al. 2008; Milone et al. 2009 and their subsequent papers) or split RGB sequences (Niederhofer et al. 2017a,b). 
Note that the eMSTOs shown in some Magellanic Cloud clusters (MCCs) can be, however, contaminated by the stellar rotation of massive stars (Bastian \& de Mink 2009; Li et al. 2012; Yang et al. 2013; Li et al. 2014).

Li et al. (2016a) found the presence of relatively younger (0.5--1 Gyr old) generation stars with clearly discrete sequences in the CMDs of in particular the
intermediate-age ($\sim$1.5 Gyr old) massive clusters NGC 1783, NGC 1806 and NGC 411 in the Magellanic Clouds.
The age differences between old- and young-population stars are 440 Myr and 520 Myr for the two younger sequences in NGC 1783, 1.02 Gyr for NGC 1806 and 1.06 Gyr for NGC 411 (Li et al. 2016a).
Pristine gas and gas ejecta from old-generation stars are expected to have been removed from the cluster efficiently at an early stage (Bastian \& Strader 2014).
Mucciarelli et al. (2008, 2014) have not found any evidence of spreads of light-element abundances in NGC 1783 and NGC 1806. 
In addition, the estimated mass ratio of the SG to FG stars in these MCCs is only $\sim$1\% (Li et al. 2016a).
Thus the origin of the multiple populations in these MCCs might be different from that of old globular clusters.
Li et al. (2016a) suggested that these SG populations may have been formed inside the clusters as a consequence of the accumulation of 
external gas. This ad hoc formation scenario is supported by For \& Bekki (2017) who discovered 
newly formed young stellar objects in young star clusters (100--400 Myr) in the LMC.
However, there is still a lack of understanding; there has been no evidence of sufficient amounts 
of gas accumulation in young massive clusters in the Magellanic Clouds (Bastian \& Strader 2014). 
In addition, the SG stars in NGC 1783, NGC 1806 and NGC 411 are less centrally concentrated 
than the FG stars in contrast with the theoretical study of the formation of the multiple stellar populations done by D'Ercole et al. (2008).

An alternative scenario can be proposed for the formation of the multiple populations in these clusters, that is, the minor merger scenario.
The merger scenario has already been proposed to explain the reversed radial distribution in globular clusters by many observational and theoretical studies (e.g., Carretta et al. 2010; Amaro-Seoane et al. 2013; Lee 2015; Gavagnin et al. 2016).
The estimated total masses of younger-sequence stars are 372 \msun$ $ and 250 \msun$ $ for NGC 1783 sequences A and B, respectively,
527 \msun$ $ for NGC 1806 and 560 \msun$ $ for NGC 411 (Li et al. 2016a).
It has been found that relatively small clusters ($10^{3}$--$10^{4}$\msun) with ages from a few hundred Myr to 1 Gyr are very abundant
in the LMC and SMC (e.g., Hunter et al. 2003; de Grijs \& Anders 2006; Glatt et al. 2010; Baumgardt et al. 2013).
Although the probability of close encounters of star clusters that lead to a merger of clusters is very low in general,
for dwarf galaxies like the Magellanic Clouds, mergers of clusters can be more probable because of the small velocity dispersion of the cluster system (van den Berg 1996; Gavagnin et al. 2016).
Binary (or multiple) star clusters observed in the Magellanic Clouds (e.g., Bica et al. 1999; Dieball et al. 2002)
are known to be a consequence of fragmentation during massive star formation.
However, the existence of binary clusters with a large age difference (Vallenari et al. 1998; Leon et al. 1999)
and the lack of binary star clusters with relatively old ages ($>$300 Myr) (Dieball et al. 2002)
may support the possibility that some binary star clusters can be the interim stage of the merger of star clusters.

We note that the Li et al. (2016a) results have led to vigorous discussion
in the literature. Cabrera-Ziri et al. (2016) claimed that the SG
sequences in these clusters may be the result of incorrect background
subtraction. We briefly addressed these concerns in Li et al. (2016b);
here we provide a more detailed rebuttal to the Cabrera-Ziri et
al. (2016) challenge.

Our arguments in support of the reality of the younger sequences
associated with our sample clusters are, in essence, fivefold:

1. The two younger sequences in NGC 1783 are both more centrally
concentrated than expected from a uniformly distributed background
field population. Their radial profiles are less steeply peaked than
that of the cluster's bulk stellar population, but all SG stars are
clearly contained within twice the cluster's core radius (as defined
by the bulk population); the distribution of the SG stars is not
consistent with a uniform field distribution. 

2. Cabrera-Ziri et al.'s (2016) main challenge to the Li et
al. (2016a) results is based on how the latter authors dealt with
background field contamination. The former authors claim that the
statistical field-star subtraction employed by Li et al. (2016a)
resulted in over-subtraction of the field contribution, thus causing
artificially enhanced younger sequences. However, Li et al. (2016a)
went to great lengths to ensure the reality and reproducibility of
their results by exploring the effects of the grid cell sizes used for the
statistical background subtraction. As shown in their Supplementary
Information, only for unrealistically large or small grid cells did
they not recover the younger features. This underscores that the
younger sequences are indeed unlikely the result of incorrectly
dealing with the statistical nature of a uniformly distributed
background population.

3. Moreover, Li et al. (2016a; their Supplementary Information)
clearly pointed out that a binary sequence runs	parallel to the	main
sequence in the eastern part of the reference field region used	by
Cabrera-Ziri et al. (2016). This suggests that this particular
reference field may be contaminated by recent star-forming activity,
since only high-mass-ratio binaries can survive in such conditions.
In fact, in private communication with the Cabrera-Ziri et al. (2016)
team, Li et al. had pointed this out to these latter authors, but this
was not taken into account in Cabrera-Ziri et al.’s (2016)
challenge. In addition, the stellar number density in the reference
field region adopted by Cabrera-Ziri et al. (2016) is significantly
higher than that in the cluster’s periphery, so	that adoption of the
entire image as reference field is inappropriate.

4. The widths of all younger sequences in Li et al. (2016a) are such
that any reasonable age spread as expected from a mixed population of
background stars cannot be accommodated. In addition, the
average age of the field populations in the LMC and SMC are of order
$10^9$ yr or older (e.g., Rubele et al. 2012, 2015); the younger
features found in our sample clusters are significantly (at least an
order of magnitude) younger than this, and so they are unlikely
associated with a uniformly mixed background field population.

5. In Li et al. (2016a) we did not only focus on the distribution of
the young(er) main-sequence stars, but we also found a continuous centrally peaked
spatial distribution defined by the more evolved stars on the NGC 1783
giant branch. Cabrera-Ziri et al. (2016) only used the main-sequence
stars to dismiss the Li et al. (2016a) results. 

While on the basis of an assessment of the background filed characteristics 
alone we cannot reach any firm conclusion as to the nature of the younger sequences,
 we believe that these five arguments taken together provide a strong
suggestion that the younger sequences associated with our sample
clusters are more likely composed of cluster members than of field
stars. Nevertheless, we realize that some level of controversy may
remain between the Li et al. (2016a,b) and Cabrera-Ziri et al. (2016)
teams. At the very least, we contend that the dismissal of the Li
et al. (2016a) results by Cabrera-Ziri et al. (2016) was not supported
by the data. As such, we believe that it is worth exploring how
intermediate-age star clusters, including NGC 411 and NGC 1806, might
have collected younger populations. That is what we set out to do in
this paper.
 
In this study, we perform $N$-body simulations based on both the ad hoc formation scenario 
that SG stars are formed inside the cluster using gas originating from the external intergalactic 
medium as suggested by Li et al. (2016a) and a minor merger scenario of unequal mass clusters with different ages. 
The main purpose of this study is not to fit these particular clusters to simulation models 
but to understand which astrophysical properties induce the observational properties of these clusters 
(e.g., the radial trends of the SG-to-FG number ratio) 
based on different scenarios and provide observable data (e.g., morphology, kinematic properties) 
that can be used for future observations to better illustrate the formation of the multiple populations in these clusters.
In Section 2, we introduce the method, astrophysical properties of the target clusters and the initial conditions for the simulations.
In Section 3, the main results are presented. We discuss what kinds of observational properties can help constrain the formation of the multiple populations in Section 4.
Our conclusion and final remarks follow in Section 5.

\section{METHODS and MODELS}

\begin{table*}
  \begin{center}
  \caption{Parameters for representative models}
  \begin{tabular}{l c c c l c c c c c c c c c}
  \hline
  \hline
  & \multicolumn{3}{c}{FG system}& & \multicolumn{5}{c}{SG system} & & \multicolumn{3}{c}{Present day$^{\ddagger}$}\\
  Cluster & $M_{0}$ & $N_{\rm tot,0}$ & $r_{\rm c,0}$$^{\dagger}$ & Model ID & $M_{0}$ & $N_{\rm tot,0}$ & $r_{\rm c,0}$ & $(x,y)$ 
  & $(v_{x},v_{y})$ & & $M$ &$r_{\rm c}$ &$M_{\rm SG}$ \\\cline{2-4}\cline{6-10}\cline{12-14}
  & (1) & (2) &(3) &(4) &(5) &(6) &(7) &(8) &(9) & &(10) &(11)&(12)\\
  \hline
  NGC 411 & $4.1\times10^4$ & 91,000 & 4.1 & A411 & 630 & 1,400 & 10 & - & - & & $2.88\times10^{4}$ & 6.13 & 359 \\
   & & & & M411 & 3,600 & 8,000 & 2.4 & (250,50) & (--1,0) & & $2.89\times10^{4}$ & 6.60 & 585 \\
  \hline
  NGC 1806 & $1.4\times10^{5}$ & 313,000 & 5.0 & A1806 & 1,350 & 3,000 & 17 & - & - & & $1.01\times10^{5}$ & 6.83 & 640 \\
  & & & & M1806 & 6,750 & 15,000 & 3.4 & (200,40) & (--1,0) & & $1.01\times10^{5}$ & 6.88 & 840 \\ 
  \hline
  \end{tabular}\\
\begin{flushleft}
  (1) Initial mass of the FG cluster (\msun), (2) Initial total number of stars for the FG cluster, (3) Initial core radius for the FG cluster (pc),
  (4) Simulation models. ``A'' and ``M'' represent the ad hoc formation and the merger model, respectively, (5) Initial mass of the SG subsystem (\msun), 
  (6) Initial total number of stars for the SG subsystem, (7) Initial core radius for the SG subsystem (pc), (8) Initial separation of FG and SG clusters (pc),
  (9) Initial relative velocity between FG and SG clusters (km s$^{-1}$), (10) present-day total mass (\msun), (11) present-day core radius (pc), 
  (12) present-day total SG mass (\msun)\\
  $^{\dagger}$ the core radii are obtained from the surface number density profiles ($\Sigma(r_{\rm c})=\Sigma_{0}/2$).\\
  $^{\ddagger}$ these properties are the average values estimated from each simulation model for different realizations. $M$ and $M_{\rm SG}$ are the total mass within $R<2r_{\rm eff}$.
  {\it Note}: the half-number radius is $\sim$1.5 ($\sim$1.6)$r_{\rm c}$ and $\sim$60\% ($\sim$58\%) of stars are within $R<r_{\rm eff}$ for the NGC 411 (1806) model.
\end{flushleft}
  \end{center}
\end{table*}

\subsection{Simulation methods}

The $N$-body simulations of star clusters in this paper were carried out with {\sc nbody6++gpu} 
(Wang et al. 2015, 2016)\footnote{DRAGON simulations, see http://silkroad.bao.ac.cn/dragon/}. 
This code is based on the well-known legacy code {\sc nbody6} and its precursors, developed by S. Aarseth since the 1960s for direct $N$-body integration 
of stellar systems (see for an overview Aarseth 1999, 2003). 
It uses a fourth-order Hermite time integrator based on two time-steps only (Makino \& Aarseth 1992),
 and the AC (Ahmad \& Cohen 1973) neighbour scheme. 
Another important feature is the accurate treatment of binary and close encounter dynamics, 
which are crucial physical processes in star clusters, by employing the algorithms of Kustaanheimo \& Stiefel (1965) and chain 
regularization (Mikkola \& Aarseth 1993). 

Spurzem (1999) presented the first massively parallel implementation of the 
code (named {\sc nbody6++}), using message passing interface (MPI) on MIMD supercomputers. Later, 
a GPU-based parallelized version ({\sc nbody6gpu}), designed for a desktop or a single computer 
node with GPU was developed by Nitadori \& Aarseth (2012). 
Wang et al. (2015, 2016) provide a combination of MPI parallelization 
with the use of many GPUs (using GPU acceleration for every MPI process),
 with the added feature of OpenMP (or SSE/AVX) used for the neighbour forces. 
This latest variant, called {\sc nbody6++gpu} can be used across 
multiple GPU-accelerated nodes on supercomputers combining MPI, OpenMP and CUDA-based 
GPU parallel computing principles. It is a code optimized for a truly hybrid architecture. 
This technical development enabled the first million-body simulations of globular clusters (Wang et al. 2016). 
All codes include the single and binary stellar evolution 
recipes described by Hurley et al. (2000, 2002), and galactic tidal fields. In addition to these standard 
recipes the natal kicks of neutron stars have been updated (Hobbs et al. 2005)
 in the treatment of velocity kicks for neutron stars (NSs) and black holes (BHs)
when they form after supernova explosions, and employing a fallback scenario according to
 Belczynski et al. (2002) for the formation of a certain mass range of stellar BHs 
(for more details see Wang et al. 2016). Note that spectral synthesis
routines (GALEV) are now coupled to our {\sc nbody} code, which will allow us also to 
present spectra and CMDs of our simulated star clusters in every desired waveband (Pang et al. 2016).

\subsection{Target clusters \& initial conditions}

The target clusters in this paper are NGC 411 in the SMC and NGC 1806 in the LMC. The total masses of NGC 411 and NGC 1806
are $3.2\times10^{4}$\msun$ $ (Li et al. 2016a) and $1.1\times10^{5}$\msun$ $ (Goudfrooij et al. 2011), respectively.
The core and effective radii for NGC 411 are 20--25 arcsec (6.0--7.5 pc) and 50 arcsec (15 pc) and those for NGC 1806 are 
$\sim$30 arcsec ($\sim$7.2 pc) and 60 arcsec (14.4 pc), respectively (Li et al. 2016a).
The ages of NGC 411 and NGC 1806 are 1.38 Gyr and 1.51 Gyr and the ages of the SG stars are 320 Myr and 500 Myr, respectively (Li et al. 2016a).
In this study, we exclude NGC 1783 from our target clusters because NGC 1783 has two younger sequences
which cause a huge degree of freedom of parameter space to be taken into account.

The initial conditions used in this study are generated by the {\sc mocca} code (Giersz et al. 2013). The radial distributions 
of the initial FG and SG systems follows a Plummer density distribution.
The initial masses of stars are given following Kroupa et al. (1993) with the mass range between 0.08\msun$ $ and 100\msun. 
We considered single and binary stellar evolution (Hurley et al. 2000, 2002) 
but did not include primordial binaries. 
No tidal field from the Milky Way or the Magellanic Clouds is taken into account in this study for simplicity.

Table 1 shows the parameters of the representative simulation models performed in this study, such as the initial number of systems, the initial core radius, the current mass and size.
In order to generate multiple population models, first we simulate the FG system only until the time when the SG stars are formed. We stop the
simulation and extract the physical data such as mass, position, velocity, stellar type and stellar evolution time-scale.
Then we combine the FG stars with the SG stars generated independently based on different formation scenarios, either ad hoc formation or the merger model, but we set a label to keep track of the generation of individual stars.
The velocities of the SG stars for the ad hoc formation model are assumed to follow the velocity dispersion of the FG stars 
because addition of a small fraction ($\sim$1\%) of diffuse SG stars rarely affects the velocity structure.
On the other hand, the velocities of the SG stars in the merger model are determined by their own velocity dispersion based on the mass distribution of the SG cluster
since the small cluster of the SG stars is initially separated from the FG cluster and forms independently.
We performed five simulations with different realizations of the initial conditions for the SG cluster for each formation model to reduce stochastic effects induced by the small number of SG stars.

Columns 9 and 10 in Table 1 show the relative positions and velocities of the centre of mass of the SG system for the merger scenario when the SG cluster is formed. 
The numbers of known clusters and associations are 6659 and 1237 in the LMC (Bica et al. 1999) and the SMC (Bica \& Dutra 2000), respectively. 
The mean distance between clusters and associations is about 300 pc and 250 pc for the LMC and the SMC, simplistically assuming that clusters and associations are roughly uniformly distributed.
The relative velocity between clusters is chosen to be smaller than the cluster internal velocity dispersion (1.8 km s$^{-1}$ for NGC 411 and 3.2 km s$^{-1}$ for NGC 1806), otherwise a merger event is not likely (Gavagnin et al. 2016). 
For the LMC, the encounter rate of clusters is $\sim$0.5--1 Gyr$^{-1}$ (Dieball et al. 2002). 
For a given cluster velocity dispersion of $\sim$30 km s$^{-1}$ (e.g., Grocholski et al. 2006), the fraction of close encounters 
that can lead to a merger ($v_{\rm rel}<$ 3.2 km s$^{-1}$) is $\sim$21\% based on the initial parameters for the M1806 model. 
For the M411 model in the SMC, all close encounters may result in a merger ($v_{\rm rel}<$ 1.8 km s$^{-1}$) with a given velocity dispersion of SMC clusters of 23.6 km s$^{-1}$ (Parisi et al. 2009). 
We choose the relative positions and velocities arbitrarily to ensure that there are one or two pericentre passages between the two clusters;
$\sim$90\% of close encounters leading to a merger have the relative velocity less than 1 km s$^{-1}$ based on the M411 model parameters ($\sim$3\% for the M1806 model).

\section{RESULT}

Figure 2 of Li et al. (2016a) clearly shows the reversed radial distribution (i.e. SG stars are less centrally concentrated than FG stars). 
First, we tested a simulation using initial conditions for NGC 411 with more centrally concentrated SG stars than FG stars, 
based on the formation of the multiple populations in globular clusters through the self-enrichment scenario (D'Ercole et al. 2008). 
However, the result did not show any evidence of a reversal of the radial distribution or spatial mixing,
and the radial trend remained similar to the initial one until the end of simulation 600 Myr after the SG stars form.
This is because the spatial mixing among different populations is mainly due to the preferential loss caused by the tidal field of the FG stars which are spatially more extended
and complete spatial mixing can be reached when the cluster loses a significant fraction of its mass (Vesperini et al. 2013; Miholics et al. 2015).
Therefore, if the SG stars formed inside the cluster from the collected gas from outside, as suggested by Li et al. (2016a), 
the SG stars should be initially less concentrated than FG stars. 

\subsection{NGC 411}

\begin{figure}
  \centering
  \includegraphics[width=84mm]{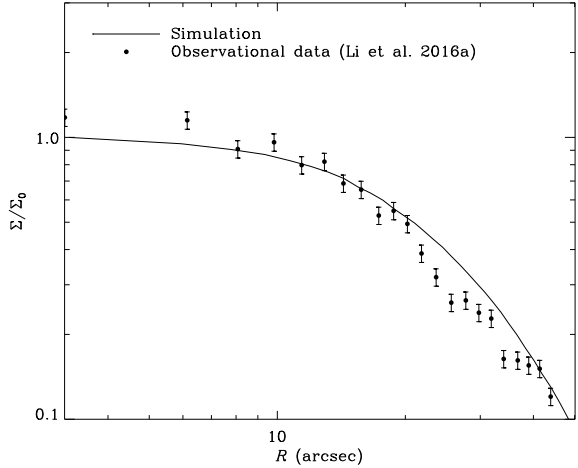}
  \caption{ Current surface number density profile of NGC 411 ($T=$1.38 Gyr). The solid line and dots represent the simulation result of the A411 model
  and the observational data from Li et al. (2016a), respectively.}
  \label{den411}
\end{figure}

In Fig. \ref{den411}, we show the radial profile of the surface number density for NGC 411 with a truncation radius of 50 arcsec. The simulation result for the A411 model and 
the observational data from Li et al. (2016a) show good agreement in general. Since the surface density profile is dominated by FG stars occupying $\sim$99\% of the total mass, 
SG stars from either the ad hoc formation or merger scenarios do not affect the overall profile.

\begin{figure}
  \centering
  \includegraphics[width=84mm]{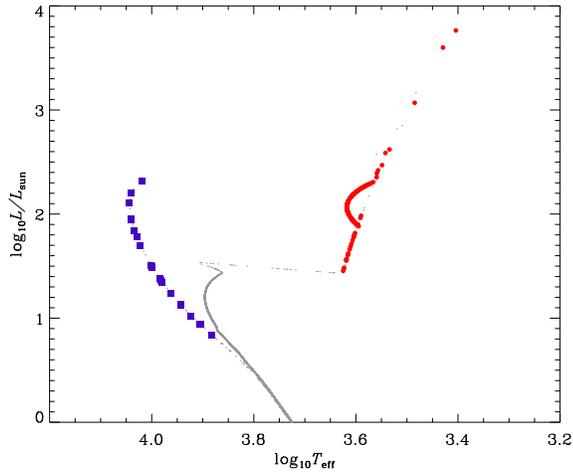}
  \caption{ Simulated Hertzsprung--Russell (HR) diagram for the A411 model at $T=$1.38 Gyr. 
  Grey dots, blue squares and red circles represent all stars,
  SG stars with log$_{10}L/$L$_{\odot}>0.8$ and FG red giant branch (RGB) and red clump (RC) stars, respectively. 
  The age of the SG stars is 320 Myr. Only stars within $2r_{\rm c}$ are plotted.}
  \label{hrd411}
\end{figure}

Fig. \ref{hrd411} shows the simulated Hertzsprung--Russell (HR) diagram for NGC 411 obtained from the A411 model. 
The $x$- and $y$-axes show the effective temperature and luminosity in solar units, respectively. 
Stars and binaries in the simulations evolve based on the stellar evolution recipe developed by Hurley et al (2000, 2002). 
There are two clear sequences of FG and SG stars with ages of 1.38 Gyr and 320 Myr, respectively. A metallicity of $Z=$0.002 is used in this model ($Z=$0.004 for NGC 1806) and there is no metallicity difference assumed between FG and SG stars for simplicity.
We did not consider stellar rotation which leads to the eMSTO in the HR diagram. 
To compare our results to the observations presented by Li et al. (2016a), we need to select FG and SG stars in the same way as Li et al. (2016a).
We choose SG stars of log$_{10}L/$L$_{\odot}>0.8$, the MSTO luminosity of FG stars. 
SG stars that are too close to red giant branch (RGB) and red clump (RC) FG stars in the HR-diagram to be distinguished
from FG stars are excluded from the selection of SG stars but counted as FG star samples as done observationally. 

\begin{figure}
  \centering
  \includegraphics[width=84mm]{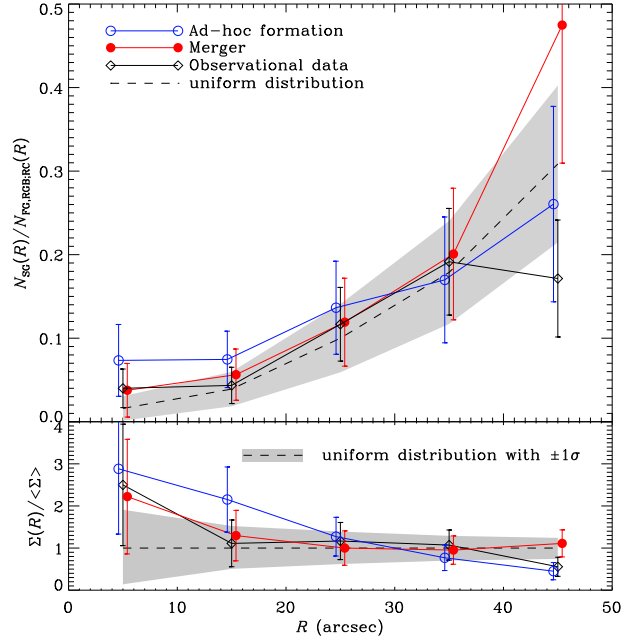}
  \caption{ {\it Top panel}: Number ratio of SG stars to FG stars (RGB and RC) in NGC 411 (see Fig. \ref{hrd411} and text for the selection of SG and FG stars). 
   Five simulations with different initial realizations of SG stars are combined for the number ratios in each simulation model.
   Error bars represent Poissonian errors. The dashed line indicates the ratio profile for a uniform distribution of SG stars, and the grey area shows the 1$\sigma$ confidence level (see the text). {\it Bottom panel}: Surface number density of SG stars normalized to the mean surface number density. The dashed line and grey area are a uniform distribution and its 1$\sigma$ confidence intervals, respectively. } 
  \label{nrp411}
\end{figure}

In Fig. \ref{nrp411} we show the radial profile of the number ratio of SG to FG stars (RGB and RC) and the surface number density profile of SG stars normalized to their mean surface number density for NGC 411, see the blue squares and red circles in Fig. \ref{hrd411} for the selection of stars.
We counted FG and SG stars within radial bins and obtained the number ratio for the simulation results and the observational data. 
To ensure that the observed SG stars are not caused by field contamination but cluster members (see Section 1), we simulated a thousand uniform distributions with the same number as the number of SG stars in the observations. 
Then we plotted the number ratio profile and the surface density profile with the 1$\sigma$ confidence level. 
Although individual data points are within the 1$\sigma$ confidence level from the uniform distribution because of the small sample size, 
 the surface density profiles from the simulation models and the observational data show a clear trend of a monotonic decrease with radius, which implies that there is a distinct underlying structure other than a uniform distribution.
Both the ad hoc formation and merger scenarios can reproduce the reversed radial distribution seen in the observations of NGC 411. 
Since the two-body relaxation time at the given radii for SG stars based on the ad hoc formation scenario is large, 
SG stars do not undergo significant dynamical evolution and retain a similar distribution to the initial one. 
The current radial distribution of SG stars closely depends on the initial conditions, such as the total mass and length scales of the SG system. 
The increase of the initial total mass and the decrease of the initial size of the SG system lead to an increase of the overall ratio profile while the decrease of the total mass and the increase of the size lead to a decrease of the profile. 
The slope of the ratio profile depends on the initial size of the SG system. Note that, however, the initially less concentrated SG stars are rather unrealistic because clusters should have a deep potential well to retain the gas against the effect of ram pressure (Conroy \& Spergel 2011) and this necessarily results in the collection of the gas at the centre by a cooling flow in order to form stars (D'Ercole et al. 2008). 

On the other hand, the SG cluster in the merger model has an initially much smaller core radius than that of the SG system for the ad hoc formation model, as shown in Table 1. 
The current radial distribution of SG stars in the the merger model is the result of the combined effect of the cluster orbits, mass and size of the merging clusters. 
Gavagnin et al. (2016) found that the radial distribution of the ratio of the two populations after the merger event 
depends on the mass ratio of the two clusters and the central density (half-mass radius) of the smaller cluster. 
They suggested that, for merged clusters, stars originating from the smaller cluster can be less centrally concentrated than stars from the larger cluster  
when $M_{12}\cdot\rho_{12}>1$, where $M_{12}$, $\rho_{12}$ are the ratio of the initial total mass and the central density of the two clusters, respectively.
From the test simulations, we confirmed that there is a rough correlation between $M_{12}\cdot\rho_{12}$ and the ratio of the half mass radii of the two populations in the merged cluster, $r_{\rm h,1}/r_{\rm h,2}=1/(M_{12}\cdot\rho_{12})+\alpha$, where $\alpha\sim0.25$.
In the cases of the clusters we are focusing on, $M_{12}\cdot\rho_{12}\sim18$ (for the M1806 model, $M_{12}\cdot\rho_{12}\sim79$). 
Thus, the merger event that could happen to NGC 411 more likely leads to the very diffuse radial distribution of SG stars and the reversed ratio profile, correspondingly.

\subsection{NGC 1806}

\begin{figure}
  \centering
  \includegraphics[width=84mm]{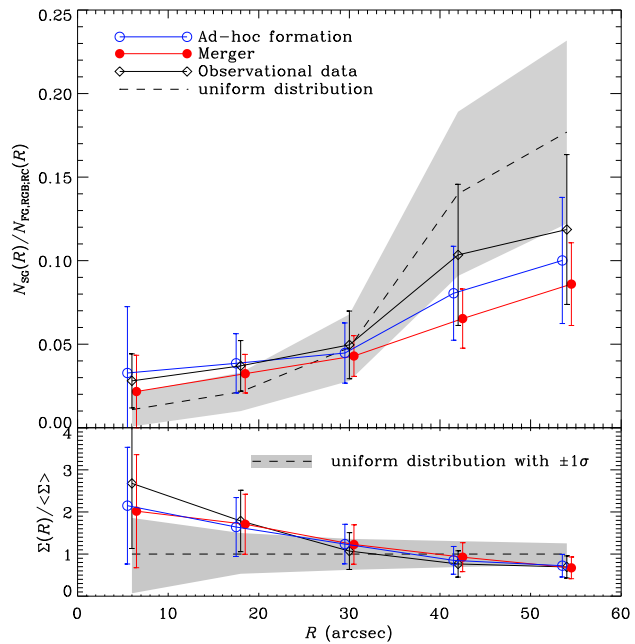}
  \caption{ Number ratio of SG stars to FG stars (top panel) and the surface number density profile (bottom panel) in NGC 1806 as Fig. \ref{nrp411}.}
  \label{nrp1806}
\end{figure}

Fig. \ref{nrp1806} shows the radial profile of the number ratio of SG stars to FG stars (RGB and RC) and the surface number density of SG stars in NGC 1806. 
SG stars with log$_{10}L/$L$_{\odot}>0.75$ are selected for this figure.
The overall ratio profiles are smaller than those for NGC 411 because the cluster total mass of NGC 1806 is about three times larger than that of NGC 411 while the total masses of the SG stars are similar.
The reversed radial trend of the SG-to-FG number ratio is well reproduced for both the ad hoc formation and merger models. 
The monotonic decrease of the surface density profiles is also obvious for NGC 1806 despite the small significance levels of the data points.

\section{DISCUSSION}

\subsection{Structural properties}

\begin{figure*}
  \centering
  \includegraphics[width=175mm]{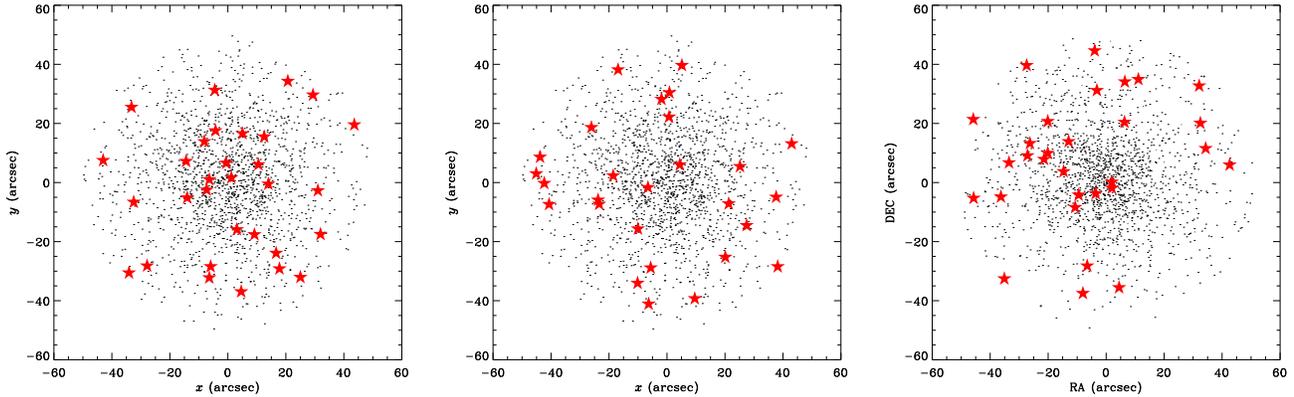}
  \caption{ Spatial distribution of stars in the $xy$ plane for A411 (left panel), M411 (middle panel) and observational data (right panel) from Li et al. (2016a).
  In the simulation results, black dots represent FG stars with log$_{10}L/$L$_{\odot}>0.7$ and red stars are SG stars satisfying the selection criteria as mentioned in Section 3.1, respectively. The numbers of SG stars are 31 (A411), 27 (M411) and 30 (Li et al. 2016a).
  For the observational data, dots are FG stars with $B<$22.5 mag and red stars are SG stars selected by Li et al. (2016a). }
  \label{xy}
\end{figure*}

To trace back to the dynamical origin of the multiple populations from the current observations of NGC 411, 
we examined the structural properties of the SG stars. 
Fig. \ref{xy} shows the spatial distribution of FG and SG stars at $T=1.38$ Gyr for the simulation models and observational data (Li et al. 2016a). 
We used SG stars in the simulations which satisfy the selection criteria described in the previous section.
By looking at the projected positions of the SG stars, we could not find structural similarity 
between the observational data and either the ad hoc formation model or the merger model.
So, we adopted the minimum spanning tree algorithm to quantify the structural properties of SG stars for both the simulations and the observational data (see also Allison et al. 2009).
We measured a dimensionless quantity $\Lambda$ defined as $\sigma_{\rm mst}/\langle l_{\rm mst}\rangle$, where $\langle l_{\rm mst}\rangle$ and $\sigma_{\rm mst}$ are 
the average and standard deviation of the length of the minimum spanning tree, respectively.
$\Lambda$ increases with the presence of substructures like cores, clumps and filaments.

\begin{figure}
  \centering
  \includegraphics[width=84mm]{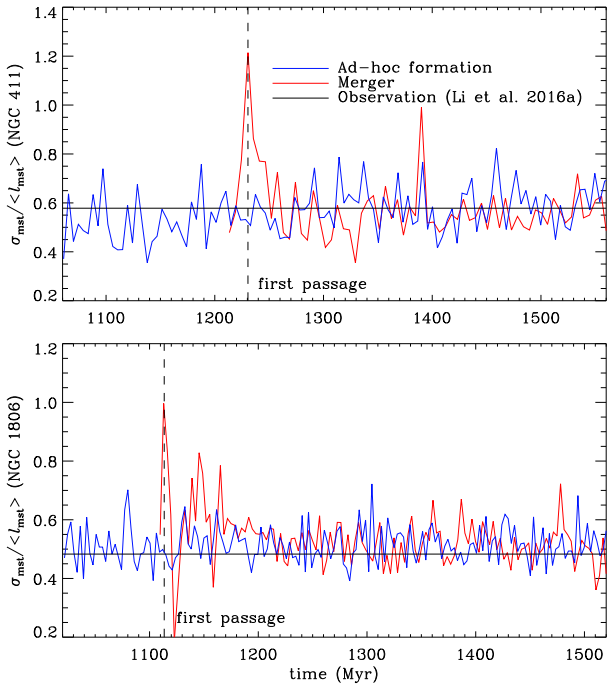}
  \caption{ Time evolution of $\Lambda$ ($\equiv\sigma_{\rm mst}/\langle l_{\rm mst}\rangle$; for more details, see the text). 
  Blue and red lines represent the evolution of $\Lambda$ for the ad hoc and merger models, respectively. 
  The vertical dashed lines indicate the moment of the first pericentre crossing of the SG system for the merger models.
  The black lines indicate the value of $\Lambda_{\rm obs}$ for NGC 411 and NGC 1806 in the observational data from Li et al. (2016a).}
  \label{mst}
\end{figure}

Fig. \ref{mst} shows the time evolution of $\Lambda$ in the ad hoc and merger models for NGC 411 (top panel) and NGC 1806 (bottom panel). 
For the A411 model, $\Lambda$ does not evolve overall with time. 
It converges to $\sim$0.57 with $\sigma\sim0.09$ in the time domain which is due to the small number of SG stars.
For the merger model, $\Lambda$ peaks at the first and second pericentre passages at 1230 and 1390 Myr.
The structural discrepancy is only significant during the short time period at the passages, but not observed at other times.
The black horizontal line shows $\Lambda_{\rm obs}=0.58$ (where $\langle l_{\rm mst}\rangle=11.0$ arcsec and $\sigma_{\rm mst}=6.33$ arcsec) 
for the observational data from Li et al. (2016a).
This value is within 1$\sigma$ of $\Lambda_{\rm A411}$ so no conclusion can be reached by only using the spatial distribution.
Note that $\Lambda\sim0.46$ for a uniform random distribution, and by sampling the same number as the number of SG stars in the observational data of NGC 411 (Li et al. 2016a) the 1$\sigma$ confidence of $\Lambda$ is $\sim$0.07. 
Thus it is clear that there is a structure of SG stars in NGC 411 with 2$\sigma$ confidence. For comparison, $\Lambda_{\rm obs}$ for the FG stars is $\sim$0.67 in the observational data of NGC 411. 
For NGC 1806, $\Lambda_{\rm M1806}$ has a peak at the first pericentre passage ($\sim$1110 Myr) and $\Lambda_{\rm A1806}$ remains at a constant of $\sim0.51\pm0.06$.
$\Lambda_{\rm obs}\sim0.48$ in the observation of NGC 1806 is within the 1$\sigma$ level of $\Lambda_{\rm A1806}$, but this is also within 1$\sigma$ confidence of the uniform distribution. 
This is consistent with Cabrera-Ziri et al. (2016) who claimed that the radial distribution of SG stars in NGC 1806 is close to that of a uniform distribution. 

\subsection{Kinematic properties}
\begin{figure}
  \centering
  \includegraphics[width=84mm]{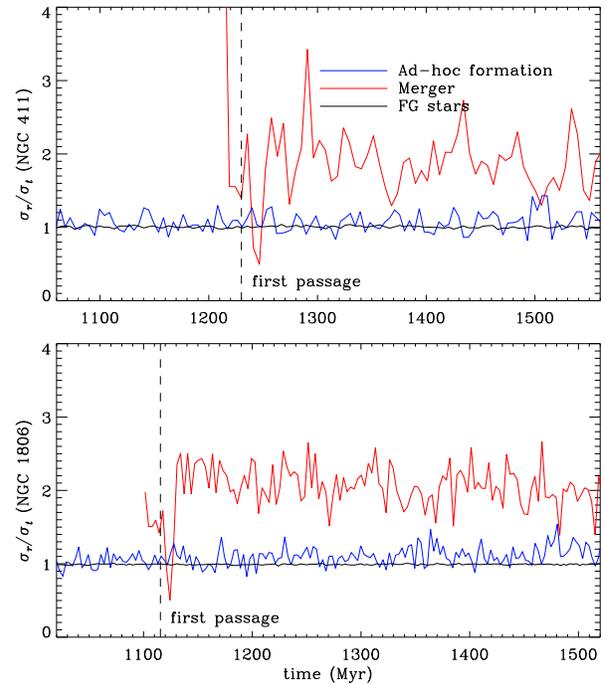}
  \caption{ Time evolution of the velocity anisotropy of FG and SG stars.
  The same stars as in Fig. \ref{xy} are used for the measurement of the velocity dispersions $\sigma_{r}$ and $\sigma_{t}$.
  The vertical dashed lines indicate the moment of the first pericentre crossing of the SG system for the merger models.}
  \label{ani411}
\end{figure}

Several observational and theoretical studies done recently for the multiple populations in globular clusters
have confirmed that SG stars show more radial anisotropy while FG stars show isotropic velocity dispersions (Richer et al. 2013; Bellini et al. 2015).
A numerical simulation done by Bellini et al. (2015) shows that the radial anisotropy of SG stars is a consequence 
of the spatial diffusion of the SG stars that formed in the central region of the FG system driven by the two-body relaxation. 
However, for our ad hoc formation model, SG stars form in the outer region where the relaxation 
time-scale is longer than the age of the SG stars, and therefore 
SG stars do not migrate to other regions and keep their initial isotropic velocity dispersion.
Fig. \ref{ani411} shows the time evolution of the velocity anisotropy defined as $\beta\equiv\sigma_{r}/\sigma_{t}$ 
where $\sigma_{r}$ and $\sigma_{t}$ are the projected radial and tangential velocity dispersions, respectively.
SG stars used for the calculation of the velocity anisotropy are the same as those in Fig. \ref{mst} 
and FG stars with log$_{10}L/$L$_{\odot}>0.5$ are used for a larger sample size. 
For the ad hoc formation model, the anisotropy parameter $\beta$ for both FG and SG stars is unity which means an isotropic velocity dispersion. 
The fluctuation of $\beta$ for the SG stars is owing to the small sample size of the SG stars. 
On the other hand, for the merger model, $\beta$ fluctuates excessively immediately after the first pericentre passage and settles to $\beta\sim1.9$ ($\beta\sim2.0$ for the M1806 model). 
This radial anisotropy is long-lasting through the end of the simulation after the merger event. 

A merger of clusters does not always lead to radial anisotropy.
If the orbits of two clusters are close to circular, the velocity anisotropy after the merger event tends to be tangential (Antonini 2014).
However, the main point is that if multiple populations in the clusters NGC 1783, NGC 1806 and NGC 411
originated from mergers of two clusters, the events should leave a kinematic fingerprint in the form of a velocity anisotropy.

Lee (2015) found multiple populations in the Galactic globular cluster, M22, which show different rotational velocities
and suggested that M22 formed through a merger of two globular clusters. 
We also estimate the degree of rotation of our SG stars, defined as the ratio of 
the mean rotational velocity to the radial velocity dispersion $\langle V_{\rm rot}\rangle/\sigma_{r}$.  
$\langle V_{\rm rot}\rangle/\sigma_{r}$ for SG stars is about unity right after the first passage and preserved during the whole evolution.
Although it can rely on the cluster orbits and can be reduced depending on the inclination of the orbital plane (see Priyatikanto et al. 2016). 
a merger event will leave a significant rotational signature among different-generation stars as Lee (2015) observed.

 Unfortunately, however, we cannot obtain the kinematic information of our target MCCs from current observational capabilities.
The proper motion accuracy based on HST observation (Bellini et al. 2015) is $\sim$0.03 mas yr$^{-1}$ which corresponds to $\sim$6.7 km s$^{-1}$ for NGC 1806 and $\sim$8.1 km s$^{-1}$ for NGC 411, according to their distance.
In addition, the recent proper motion study done by the ground-based VISTA telescope (Cioni et al. 2016) measured the proper motion of the SMC stars with a proper motion accuracy of 0.07 mas yr$^{-1}$.
On the other hand, the 1-D velocity dispersions of simulation models for NGC 411 are $\sim$1.8 km s$^{-1}$ at the centre and $\sim$1.2 km s$^{-1}$ at the half-mass radius, 
and $\sim$3.9 km s$^{-1}$ at the centre and $\sim$2.7 km s$^{-1}$ at the half-mass radius for NGC 1806, respectively. 

\begin{figure*}
  \centering
  \includegraphics[width=120mm]{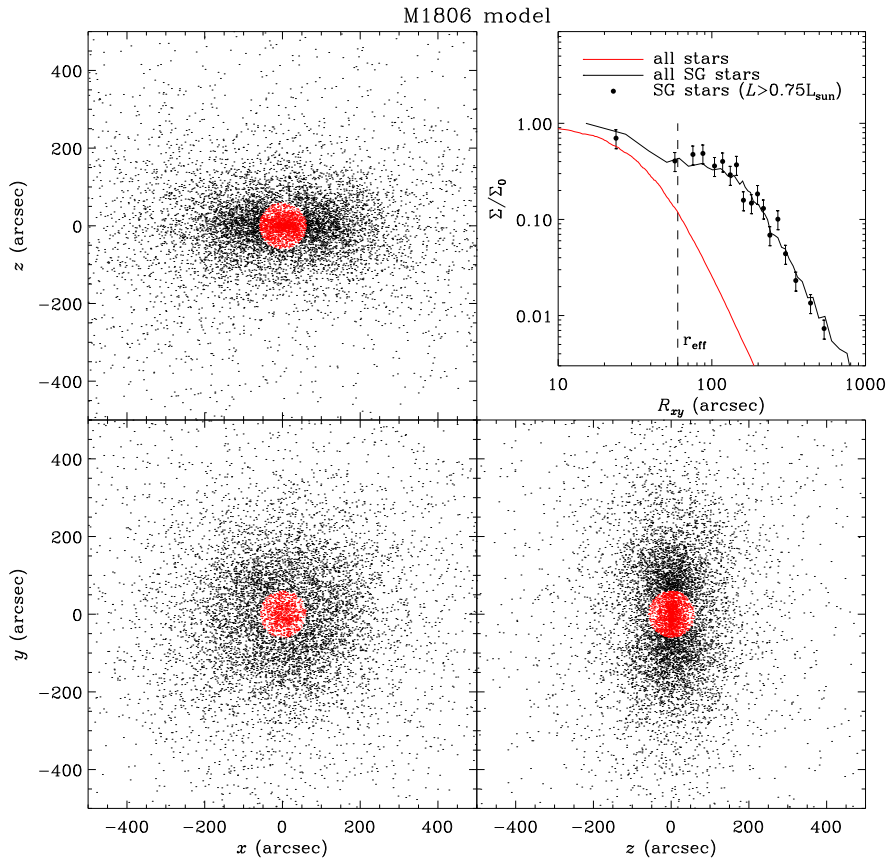}
  \caption{ Spatial distribution of SG stars in $xy$, $xz$ and $zy$ plane in the M1806 model for larger area ($R<$ 500 arcsec) at $T=1.5$ Gyr. The area with red dots indicates the region within the effective radius (60 arcsec). Top right panel shows the surface number density profiles for all star (red line), all SG stars (black line) and SG stars satisfying the selection criteria (dots with error bars) described in Section 3.2, respectively. These profiles are normalized to their central values. }
\label{memb}
\end{figure*}

\subsection{SG stars across a larger area and their membership}

Cabrera-Ziri et al. (2016) have suggested that the younger sequences in NGC 1783, NGC 1806 and NGC 411 observed by Li et al. (2016a) are the result of incorrect subtraction of background stars. 
There are two main pieces of evidence to support their claims: (1) young (SG) stars are less centrally concentrated than old (FG) stars or even show a flat distribution (e.g., NGC 1806). 
(2) there is a group of stars in the field similar to the younger-sequence stars in the observed regions of the clusters. 
However, both observational facts can be explained by a merger scenario.
According to Gavagnin et al. (2016), the radial distribution of different population stars in merged clusters is proportional to 
$M_{12}\cdot\rho_{12}$. 
Thus a merger of clusters with an extreme mass ratio leads to very diffuse distribution of the stars from the smaller cluster.
If the younger-sequence stars in NGC 1806 and NGC 411 are the result of a merger event with the large mass ratio as presented in this study, their distributions are necessarily more diffuse. It is apparent that the distribution of SG stars seems to be uniform
when $M_{12}\cdot\rho_{12}$ is so large that the core radius of the SG system is larger than the field of view.

Fig. \ref{memb} shows the spatial distribution of SG stars in the M1806 model with a larger area in the $xy$, $xz$ and $zy$ plane. 
The SG stars within $r_{\rm eff}$ are indicated by red dots. This figure clearly shows a very diffuse distribution of SG stars compared to the small cluster region. 
As can be seen from Table 1, the total mass of SG stars within 2$r_{\rm eff}$ is only $\sim$12\% of the initial mass. 
Considering the mass loss through stellar evolution, $\sim$85\% of the mass of the SG stars is located outside 2$r_{\rm eff}$. 
The top right panel shows the surface number density profiles normalized to the central value. 
Many SG stars are beyond $r_{\rm eff}$ and the ratio of the half mass radius of SG stars to FG stars is about $\sim$4.
For instance, the field region used in Cabrera-Ziri et al. (2016) for comparison to the cluster region is $\sim$3--5$r_{\rm eff}$ away from the cluster centre.
This region can be populated by SG stars with 1/5--1/10 of the central surface number density by assuming the profile of the M1806 model. 
Thus the younger-sequence stars in NGC 1783, NGC 1806 and NGC 411 can be cluster members satisfying the claims presented by Cabrera-Ziri et al. (2016) if these clusters underwent merger(s) with extreme mass ratios. 
A mosaic of observations covering a large area near the target clusters can confirm this interpretation by showing the gradient of the surface number density of larger distances or detecting merging features like tails.

\subsection{The case of NGC 1783}
For NGC 1783 which has two younger-generation star sequences, we did not perform simulations in this study because of the enormous degree of freedom of the model parameters.
There might be evidence supporting the merger scenario in the observational radial profile of the SG-to-FG number ratio from figure 2 of Li et al. (2016a)
although we caution that our observational results may be affected by small-number statistics.
Sequence B in this figure shows a smoother radial profile than sequence A, which is younger than sequence B.
In our simulations, we see that the substructure of SG such as clumps and filaments induced by the merger event can be smoothed out over time.
It roughly took $\sim$200 Myr after the first pericentre passage of the SG cluster (i.e., roughly after the second passage) for the M411 model
but this depends on the orbital parameters and cluster structure.
The radial profile of the number ratio for sequence A in NGC 1783, which is not monotonic with radius, might indicate
that the sequence A population merged with NGC 1783 recently and did not have enough time to spread into the cluster NGC 1783
if multiple populations in NGC 1783 originated from minor mergers of clusters.
Sequences A and B have $\Lambda_{\rm obs}\sim0.68$ and $\Lambda_{\rm obs}\sim0.49$, respectively. 
This also shows the presence of significant sub-structure of the sequence A population and a diffuse distribution of the sequence B population.

\subsection{Other scenarios}
There are several studies explaining the reversed radial trend through mass segregation (e.g., Larsen et al. 2015; Lim et al. 2016).
Larsen et al. (2015) distinguished three different sequences of lower RGB stars in M15 by their N abundance
and found that N-enriched RGB stars are less centrally concentrated than normal RGB stars.
However, different populations in Larsen et al. (2015) have very small age differences and
younger-generation stars have smaller masses than older-generation stars because of He enhancement.
On the other hand, SG stars selected in NGC 411 and NGC 1806 in this study are more massive that FG stars and these clusters are not old enough
in units of their half-mass relaxation time-scale to experience mass segregation ($\tau_{r,\rm c}\sim1.2$ Gyr and $\tau_{r,\rm h}\sim4$ Gyr at the time when SG stars form for the A411 model).
SG stars in NGC 411 from the observational data are even less centrally concentrated than lower-mass FG stars ($B>25$ mag).
Therefore, we can rule out the mass-segregation scenario for the explanation of the formation of the multiple populations in NGC 411 and NGC 1806.

\section{SUMMARY}

We have presented the results of direct $N$-body simulations using the {\sc nbody6++} code to understand the origin of the multiple populations in the intermediate-age MCCs NGC 411 and NGC 1806 which were recently observed by Li et al. (2016a). 
In addition to the ad hoc formation scenario where SG stars form inside the cluster using the accumulated gas from the intergalactic medium as suggested by Li et al. (2016a), 
we have taken into account a scenario of minor mergers of unequal mass ($M_{\rm SG}/M_{\rm FG} \sim 5$--10\%) clusters.

On the assumption that the younger sequences are composed of cluster members, which was justified in this paper, the most interesting feature of the multiple populations in these clusters in the observations is that SG stars are less centrally concentrated than FG stars 
in contrast with the standard theoretical predictions for the formation of the multiple populations in globular clusters (e.g., D'Ercole et al. 2008). 
It is unlikely that this reversed radial distribution has been evolved from an initially more centrally concentrated SG system 
because these clusters are expected to not have experienced significant mass loss through tidal effects, which is necessary for spatial mixing (e.g. Vesperini et al. 2013; Miholics et al. 2015). Thus, the ad hoc scenario is rather unrealistic based on the current formation models of multiple populations.

Both ad hoc formation with initially less centrally concentrated SG stars and a minor merger scenario reproduce the reversed radial distribution shown in the observations.
We found that there is no significant discrepancy in spatial distribution and the morphology of SG stars between the results from the ad hoc formation and merger scenarios.
The significance of structural differences in the merger model is only visible during the short period of pericentre passages of the clusters. 
However, the merger model shows a significant distinction in kinematic properties such as velocity anisotropy and rotation rates compared to the ad hoc formation model.
Thus, future observations aimed at obtaining kinematic properties for these clusters, as Bellini et al. (2015) did for the Galactic globular cluster NGC 2808, will help constrain the formation scenarios of the multiple populations in NGC 411, NGC 1806 and NGC 1783. 

As regards concerns related to the membership of SG stars in our target clusters (e.g., Cabrera-Ziri et al. 2016), we addressed this issue in more details; the younger sequences are less centrally concentrated than the main population but more concentrated than the uniform distribution. The widths and ages of younger sequences are significantly different with mixed background field populations.
We also investigated the spatial distributions of SG stars in the simulations and observations by obtaining the surface density profiles and by performing a minimum spanning tree analysis. 
The monotonically decreasing surface density profiles and/or the results of a minimum spanning tree analysis over $>$1$\sigma$ confidence level
indicate that the spatial distributions of SG stars are distinguished from a uniform distribution.
Moreover, the analysis done by Cabrera-Ziri et al. (2016) could not completely rule out the merger scenario with an extreme mass ratio since an extreme mass ratio merger can result in the flat radial distribution of SG stars in the inner regions and more diffuse distribution of SG stars at the outskirt of the clusters.

Note that the main purpose of this study is not to reproduce these particular clusters exactly
but to understand which astrophysical properties induce the observational properties of these clusters. 
The models presented in this study are very specific and may not be unique and exact models for the observed clusters. 
Thus, more detailed studies of a wider parameter space and more realistic astrophysical assumptions, which are not covered here, such as metallicity differences, tidal effects from the Milky Way and/or the LMC/SMC and orbital properties are necessary to obtain more precise interpretations for these clusters.

\section*{Acknowledgments} 
JH acknowledges support from the China Postdoctoral Science Foundation, Grant No. 2017M610694.
RdG acknowledges National Natural Science Foundation of China (NSFC) funding support through grants U1631102, 11373010, 11633005.
AA acknowledges support from the Polish National Science Centre through grant UMO-2015/17/N/ST9/02573.
PB acknowledges special support from the National Academy of Science of Ukraine under the Main Astronomical Observatory GRID/GPU computing cluster project. 
CL is supported by the Macquarie Research Fellowship Scheme. 
MBNK was supported by the NSFC (grants 11010237, 11050110414, 11173004, and 11573004). This research was supported by the Research Development Fund (grant RDF-16-01-16) of Xi'an Jiaotong-Liverpool University (XJTLU).
RS has been supported by the Alexander-von-Humboldt Polish Honorary Research Fellowship of the Foundation for Polish Science, and by the NSFC, grant 11673032. 
We acknowledge support from the Chinese Academy of Sciences through the Silk Road Project at the National Astronomical Observatories, Chinese Academy of Sciences (NAOC), through the Chinese
Academy of Sciences Visiting Professorship for Senior International Scientists, Grant Number 2009S1-5 (RS), and through the `Qianren' special foreign experts programme of China. 
The special GPU accelerated supercomputer laohu at the Center of Information and Computing at NAOC, funded by Ministry of Finance of People’s Republic of China under grant ZDYZ2008-2 has been
 used for the computer simulations. We acknowledge support from the Strategic Priority Research Program (Pilot B) ``Multi-wavelength gravitational wave universe'' of the Chinese Academy of Sciences (No. XDB23040100). %

\label{lastpage}
\end{document}